\documentclass[preprint, superscriptaddress, preprintnumbers,amsmath,amssymb]{revtex4}

\usepackage{booktabs}
\usepackage{color,graphicx}
\usepackage{amsmath}
\usepackage{dcolumn}
\usepackage{bm}                  % bold math
\usepackage{soul}
\usepackage{enumerate}
\usepackage{multirow}
\usepackage{mathrsfs}
\usepackage{braket}
\usepackage{longtable}
\usepackage{pdflscape}

\usepackage[dvipdfm,
            pdfstartview=FitH,
            CJKbookmarks=true,
            bookmarksnumbered=true,
            bookmarksopen=true,
            colorlinks,
            pdfborder=001,
            linkcolor=blue,
            anchorcolor=blue,
            citecolor=blue
            ]{hyperref}

\usepackage{float}

\begin{document}
%\begin{CJK*}{GBK}{song}

\title{Predictive power for superheavy nuclear mass and possible stability beyond the neutron drip line in deformed relativistic Hartree-Bogoliubov theory in continuum}

\author{Kaiyuan Zhang}
\affiliation{State Key Laboratory of Nuclear Physics and Technology,
School of Physics, Peking University, Beijing 100871, China}

\author{Xiaotao He}
\affiliation{Department of Nuclear Science and Technology, Nanjing University of Aeronautics and Astronautics, Nanjing 210016, China}

\author{Jie Meng} \email{mengj@pku.edu.cn}
\affiliation{State Key Laboratory of Nuclear Physics and Technology, School of Physics, Peking University, Beijing 100871, China}

\author{Cong Pan}
\affiliation{State Key Laboratory of Nuclear Physics and Technology,
School of Physics, Peking University, Beijing 100871, China}

\author{Caiwan Shen}
\affiliation{School of Science, Huzhou University, Huzhou 313000, China}

\author{Chen Wang}
\affiliation{Department of Nuclear Science and Technology, Nanjing University of Aeronautics and Astronautics, Nanjing 210016, China}

\author{Shuangquan Zhang}
\affiliation{State Key Laboratory of Nuclear Physics and Technology,
School of Physics, Peking University, Beijing 100871, China}

\begin{abstract}
  The predictive power of the deformed relativistic Hartree-Bogoliubov theory in continuum (DRHBc) for nuclear mass is examined in the superheavy region, $102 \le Z \le 120$.
  The accuracy of predicting the 10 (56) measured (measured and empirical) masses is $0.635$ ($0.642$) MeV, in comparison with $0.515$ ($1.360$) MeV by WS4 and $0.910$ ($2.831$) MeV by FRDM.
  Possible stability against multineutron emission beyond the two-neutron drip line is explored by the DRHBc theory, which takes into account simultaneously the deformation effects, the pairing correlations, and the continuum effects.
  Nuclei stable against two- and multineutron emissions beyond the two-neutron drip line are predicted in $_{106}$Sg, $_{108}$Hs, $_{110}$Ds, and $_{112}$Cn isotopic chains, forming a peninsula of stability adjacent to the nuclear mainland.
  This stability is mainly due to the deformation which significantly affects the shell structure around the Fermi surface.
  The pairing correlations and continuum influence the stability peninsula in a self-consistent way.
\end{abstract}

\date{\today}

\maketitle

%%%%%%%%%%%%%%%%%%%%%%%%%%%%%%%%%%%%%%%%%%%%%%%%%%%%%%%%%%
%                    begin  introduction
%%%%%%%%%%%%%%%%%%%%%%%%%%%%%%%%%%%%%%%%%%%%%%%%%%%%%%%%%%

%----------------------------------------------------------------------------------------
%\section{Introduction}
%----------------------------------------------------------------------------------------

The limit of the nuclear landscape is an interesting question to nuclear physicists~\cite{Thoennessen2013RPP,Erler2012Nature,Xia2018ADNDT}.
From light to heavy nuclei in the nuclear chart, the valley of stability changes from nuclei with a neutron number $N$ close to the proton number $Z$ to nuclei with more neutrons, due to the competition between the symmetry energy and the Coulomb interaction.
Away from the valley of stability, the lifetimes of the nuclei decrease, and finally the drip lines are reached, where the nuclei become unstable against nucleon emission.

Nuclei with even $N$ or even $Z$ are usually more stable than their odd neighbors owing to the pairing correlations~\cite{Bohr1969Book,Peter1980Book}.
As a result, the nuclear landscape defined by a one-neutron drip line with one-neutron separation energy $S_{1n}=0$ is different with one defined by the two-neutron drip line with two-neutron separation energy $S_{2n}=0$.
A typical example is shown in the lithium isotopic chain, in which $^{9}$Li is the last bound nucleus defined by $S_{1n}$ and $^{11}$Li is the last bound nucleus defined by $S_{2n}$~\cite{Tanihata1985PRL}.
With increasing $Z$, more bound nuclei with even $N$ can appear beyond the one-neutron drip line.
An impressive example is shown in \emph{giant halo} nuclei~\cite{Meng1998PRL}, in which more than 10 nuclei with even $N$ exist beyond the one-neutron drip line.

The giant halo was first predicted by the relativistic continuum Hartree-Bogoliubov (RCHB) theory in Zr isotopes~\cite{Meng1998PRL}.
Later, this prediction was supported by both non-relativistic and relativistic density functional theory studies for neutron-rich Zr and Ca isotopes~\cite{Meng2002PRC(R),Sandulescu2003PRC,Zhang2003SciChina,Terasaki2006PRC,Grasso2006PRC,Zhang2012PRC}.
The RCHB theory~\cite{Meng1996PRL,Meng1998NPA} takes into account self-consistently the pairing correlations and continuum effects and has achieved great success in describing and predicting various nuclear phenomena~\cite{Meng1996PRL,Meng1998PRL,Meng1998PLB,Meng1998Phys.Rev.C628,Meng1999Phys.Rev.C154,Meng2002PLB,Meng2002PRC(R),Zhang2005NPA,Zhang2016CPC,Lim2016PRC}.
The first nuclear mass table including continuum effects has been constructed based on the RCHB theory, which shows that the inclusion of continuum effects remarkably extends the existing nuclear landscape~\cite{Xia2018ADNDT}.
Inheriting the advantages of the RCHB theory and further including the deformation degrees of freedom, the deformed relativistic Hartree-Bogoliubov theory in continuum (DRHBc)~\cite{Zhou2010PRC(R),Li2012PRC} has been successful in the past decade~\cite{Chen2012PRC,Li2012CPL,Sun2018PLB,Zhang2019PRC,Pan2019IJMPE,Sun2020NPA,In2021IJMPE}.
Very recently, the $1s_{1/2}$ component contribution and the halo features in the $^{17}$B nucleus have been explained by the DRHBc theory~\cite{Yang2021PRL}.
The strategy and techniques to construct a mass table with the deformation and continuum effects based on the DRHBc theory have been discussed in Ref.~\cite{Zhang2020PRC}.

It would be interesting to explore the stability against neutron emission beyond the two-neutron drip line by the DRHBc theory.
We note some previous density functional theory studies on the neutron emission stability beyond the two-neutron drip line.
Several nuclei stable against two-neutron emission beyond the two-neutron drip line around the regions of $Z \approx 60, 70,$ and $100$ have been found by the nonrelativistic Skyrme Hartree-Fock-Bogoliubov calculations, which was suggested to be due to the presence of shell effects at neutron closures~\cite{Stoitsov2003PRC,Erler2012Nature}.
Similar phenomena have also been predicted by relativistic Hartree-Bogoliubov (RHB) calculations, which was attributed to the local changes in the shell structure induced by deformation~\cite{Afanasjev2013PLB}.
These investigations employed either the harmonic oscillator (HO) basis or the transformed HO basis, which is unsuitable for describing weakly bound systems with diffuse spatial density distributions~\cite{Zhou2000CPL,Zhang2013PRC}.
Stabilities against two-neutron emission beyond the two-neutron drip line were also discussed by Skyrme Hartree-Fock + BCS calculations~\cite{Gridnev2006IJMPE,Greiner2010NPA,Tarasov2013IJMPE}, in which the BCS theory is incapable of describing the pairing correlations and continuum in exotic nuclei~\cite{Dobaczewski1984NPA,Meng2006PPNP,Meng2015JPG,Meng2016Book}.
Moreover, it is unclear whether or not these nuclei stable against two-neutron emission are bound.
In other words, stability against multineutron emission beyond the two-neutron drip line was not discussed.

In this Letter we study the stability against multineutron emission beyond the two-neutron drip line by considering simultaneously the deformation effects, the pairing correlations, and the continuum effects, with the state-of-the-art DRHBc theory.

Superheavy element synthesis and the search for the ``island of stability" are at the forefront in nuclear physics.
Until now, superheavy elements, with $Z\le 118$, have been synthesized~\cite{Hofmann2000RMP,Morita2015NPA,Oganessian2017PhysScr}.
Although the island has still not been located~\cite{Sobiczewski2007PPNP}, more and more data are available now in the superheavy region, which are very useful to examine the predictive power of theoretical models.
We first examine the predictive power of the DRHBc theory for nuclear mass in the superheavy region, $102 \le Z \le 120$, by comparison with the available data~\cite{AME2020(1),AME2020(2),AME2020(3)} as well as the popular macroscopic-microscopic mass models WS4~\cite{Wang2014PLB} and FRDM(2012)~\cite{Moller2016ADNDT}.
Then based on the DRHBc calculations, the possible stabilities against two- and multineutron emission beyond the two-neutron drip line are explored.

%%%%%%%%%%%%%%%%%%%%%%%%%%%%%%%%%%%%%%%%%%%%%%%%%%%%%%%%%%
%                    begin  theoretical framework
%%%%%%%%%%%%%%%%%%%%%%%%%%%%%%%%%%%%%%%%%%%%%%%%%%%%%%%%%%

%----------------------------------------------------------------------------------------
%\section{Theoretical framework} \label{theory}
%----------------------------------------------------------------------------------------

The details of the DRHBc theory with meson-exchange and point-coupling density functionals can be found in Refs.~\cite{Li2012PRC} and \cite{Zhang2020PRC} respectively. In the DRHBc theory, the RHB equation reads~\cite{Kucharek1991ZPA},
\begin{equation}\label{RHB}
\left(\begin{matrix}
h_D-\lambda & \Delta \\
-\Delta^* &-h_D^*+\lambda
\end{matrix}\right)\left(\begin{matrix}
U_k\\
V_k
\end{matrix}\right)=E_k\left(\begin{matrix}
U_k\\
V_k
\end{matrix}\right),
\end{equation}
where $h_D$ is the Dirac Hamiltonian, $\lambda$ is the Fermi energy, and $E_k$ and $(U_k, V_k)^{\rm T}$ are the quasiparticle energy and wave function. The pairing potential is,
\begin{equation}\label{Delta}
\Delta(\bm r_1,\bm r_2) = V^{\mathrm{pp}}(\bm r_1,\bm r_2)\kappa(\bm r_1,\bm r_2),
\end{equation}
with a density-dependent force of zero range,
\begin{equation}\label{pair}
V^{\mathrm{pp}}(\bm r_1,\bm r_2)= V_0 \frac{1}{2}(1-P^\sigma)\delta(\bm r_1-\bm r_2)\left(1-\frac{\rho(\bm r_1)}{\rho_{\mathrm{sat}}}\right),
\end{equation}
and the pairing tensor $\kappa(\bm r_1,\bm r_2)$~\cite{Peter1980Book}.

The RHB equation, (\ref{RHB}), is solved in a Dirac Woods-Saxon basis~\cite{Zhou2003PRC} which can describe the large spatial extension of halo nuclei. In the Dirac Hamiltonian,
\begin{equation}
h_D(\bm{r})=\bm{\alpha}\cdot\bm{p}+V(\bm{r})+\beta[M+S(\bm{r})],
\end{equation}
the scalar and vector potentials are expanded in terms of the Legendre polynomials,
\begin{equation}\label{legendre}
f(\bm r)=\sum_\lambda f_\lambda(r)P_\lambda(\cos\theta),~~\lambda=0,2,4,\cdots,
\end{equation}
so are the pairing potential and various densities in the DRHBc theory.

%%%%%%%%%%%%%%%%%%%%%%%%%%%%%%%%%%%%%%%%%%%%%%%%%%%%%%%%%%
%                    begin  numerical details
%%%%%%%%%%%%%%%%%%%%%%%%%%%%%%%%%%%%%%%%%%%%%%%%%%%%%%%%%%

%----------------------------------------------------------------------------------------
%\section{Numerical details}\label{numerical}
%----------------------------------------------------------------------------------------

The present calculations are carried out with the density functional PC-PK1~\cite{Zhao2010PRC}.
The pairing strength $V_0=-325.0~\mathrm{MeV~fm}^3$ and the saturation density $\rho_{\mathrm{sat}}=0.152~\mathrm{fm}^{-3}$ in Eq.~(\ref{pair}) together with a pairing window of $100$ MeV.
The energy cutoff $E^+_{\mathrm{cut}}=300$ MeV and the angular momentum cutoff $J_{\max}=23/2~\hbar$ are adopted for the Dirac Woods-Saxon basis.
The above numerical details are the same as those suggested in Ref.~\cite{Zhang2020PRC} for the DRHBc mass table calculations.
For superheavy nuclei here, the Legendre expansion truncation in Eq.~(\ref{legendre}) is chosen as $\lambda_{\max}=10$~\cite{Pan2019IJMPE}.

%%%%%%%%%%%%%%%%%%%%%%%%%%%%%%%%%%%%%%%%%%%%%%%%%%%%%%%%%%
%                    begin  results and discussions
%%%%%%%%%%%%%%%%%%%%%%%%%%%%%%%%%%%%%%%%%%%%%%%%%%%%%%%%%%

%----------------------------------------------------------------------------------------
%\section{Results and Discussion}\label{results}

%----------------------------------------------------------------------------------------
\begin{figure*}[htbp]
  \centering
  \includegraphics[width=0.8\textwidth]{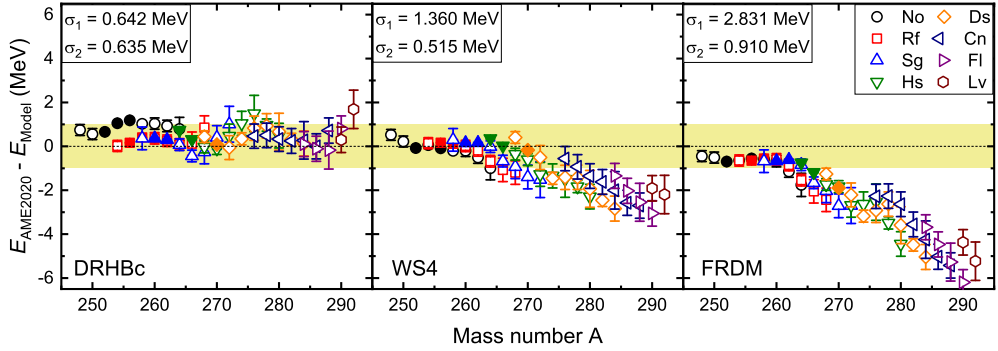}
  \caption{Mass difference between the AME2020 data~\cite{AME2020(3)} and (a) the DRHBc calculation, (b) the WS4 mass model~\cite{Wang2014PLB}, and (c) the FRDM(2012) mass model~\cite{Moller2016ADNDT} for superheavy even-even nuclei with data available ($102\le Z\le 116$). $\sigma_1$ ($\sigma_2$) represents the root-mean-square deviation from the available 56 (10) measured and empirical (measured only) masses denoted by filled and open (filled) symbols.
  }
\label{fig1}
\end{figure*}
%----------------------------------------------------------------------------------------

Figure~\ref{fig1} shows the mass difference between the DRHBc calculations and available data from the latest ``Atomic Mass Evaluation" (AME2020)~\cite{AME2020(3)} for $102\le Z\le 120$, in comparison with those for the popular macroscopic-microscopic mass models WS4~\cite{Wang2014PLB} and FRDM(2012)~\cite{Moller2016ADNDT}.
For the $10$ measured masses, the root-mean-square (rms) deviation in the DRHBc calculations is $0.635$ MeV, comparable with the $0.515$ MeV by WS4 and $0.910$ MeV by FRDM.
After including the 46 empirical data, the rms deviation in the DRHBc calculations changes slightly, to $0.642$ MeV.
However, the rms deviation for WS4 and FRDM increases, respectively, to $1.360$ and $2.831$ MeV.
The ``Atomic Mass Evaluation" empirical data are estimated from the trends in mass surface, together with all available experimental information~\cite{AME2020(2)}.
The accuracy of these empirical data is usually proved by subsequent experiments~\cite{Michimasa2020PRL,Fu2020PRC}.
In Fig.~\ref{fig1}, the DRHBc calculations provide not only a good description for the isospin dependence of the nuclear masses along the isotopic chains, but also a consistent accuracy with increasing $Z$.
These features are in stark contrast to the macroscopic-microscopic models, and demonstrate clearly the predictive power of the present calculation.

%----------------------------------------------------------------------------------------
\begin{figure*}[htbp]
  \centering
  \includegraphics[width=0.8\textwidth]{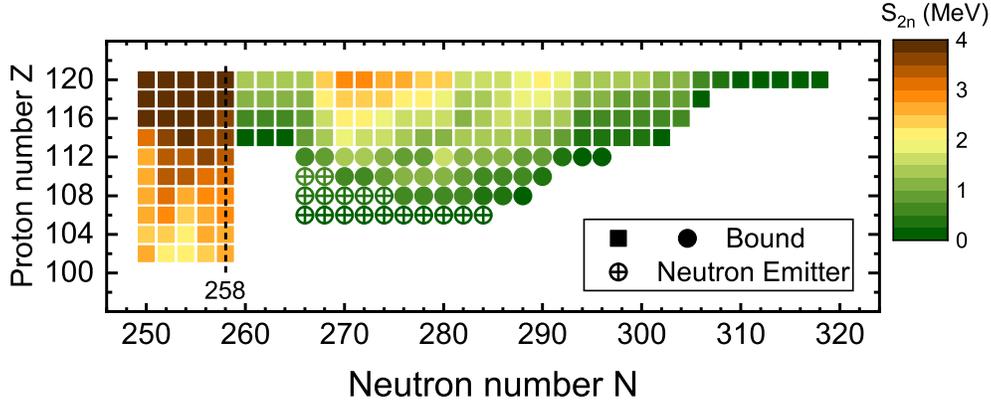}
  \caption{Two-neutron separation energy $S_{2n}$ for superheavy even-even nuclei with $102\le Z\le 120$ near the two-neutron drip line as a function of the neutron number $N$ and the proton number $Z$, where the filled squares and circles represent bound nuclei within and beyond the drip line, respectively, and the crossed circles represent nuclei stable against two-neutron emission but unstable against multineutron emission (see text for details). The dashed  line represents the possible magic number $N = 258$.}
\label{fig2}
\end{figure*}
%----------------------------------------------------------------------------------------

To explore the possible neutron emission stability beyond the two-neutron drip line, the two-neutron separation energy $S_{2n}$ for superheavy nuclei with $102\le Z\le 120$ near the two-neutron drip line is shown in Fig.~\ref{fig2}.
The sudden decrease in $S_{2n}$ at $N=258$ suggests it as a possible magic number, which agrees with previous predictions~\cite{Zhang2005NPA,Afanasjev2013PLB,Li2014PLB}.
For $_{102}$No and $_{104}$Rf isotopic chains, the existence of bound nuclei ends at $N=258$.
For the $_{106}$Sg isotopic chain, although nuclei with $266\le N \le 284$ have small positive $S_{2n}$, showing their stability against two-neutron emission, their binding energies are smaller than that of the drip-line nucleus $^{364}$Sg$_{258}$, showing their instability against multineutron emission.
For $_{108}$Hs and $_{110}$Ds isotopic chains, the stability against two-neutron emission also starts at $N=266$, and the nuclei stable against multineutron emission appear in the regions of $276\le N\le 288$ and $270\le N\le 290$, respectively.
For the $_{112}$Cn isotopic chain, all nuclei with positive $S_{2n}$ beyond the drip line are stable against multineutron emission.
For $114\le Z\le 120$ isotopic chains, $N=258$ is no longer the two-neutron drip line, which extends farther to the neutron-rich side.
Interestingly, it can be seen that a peninsula of stability for $106\le Z\le 112$ emerges from the nuclear mainland.

%----------------------------------------------------------------------------------------
\begin{figure}[htbp]
  \centering
  \includegraphics[width=0.35\textwidth]{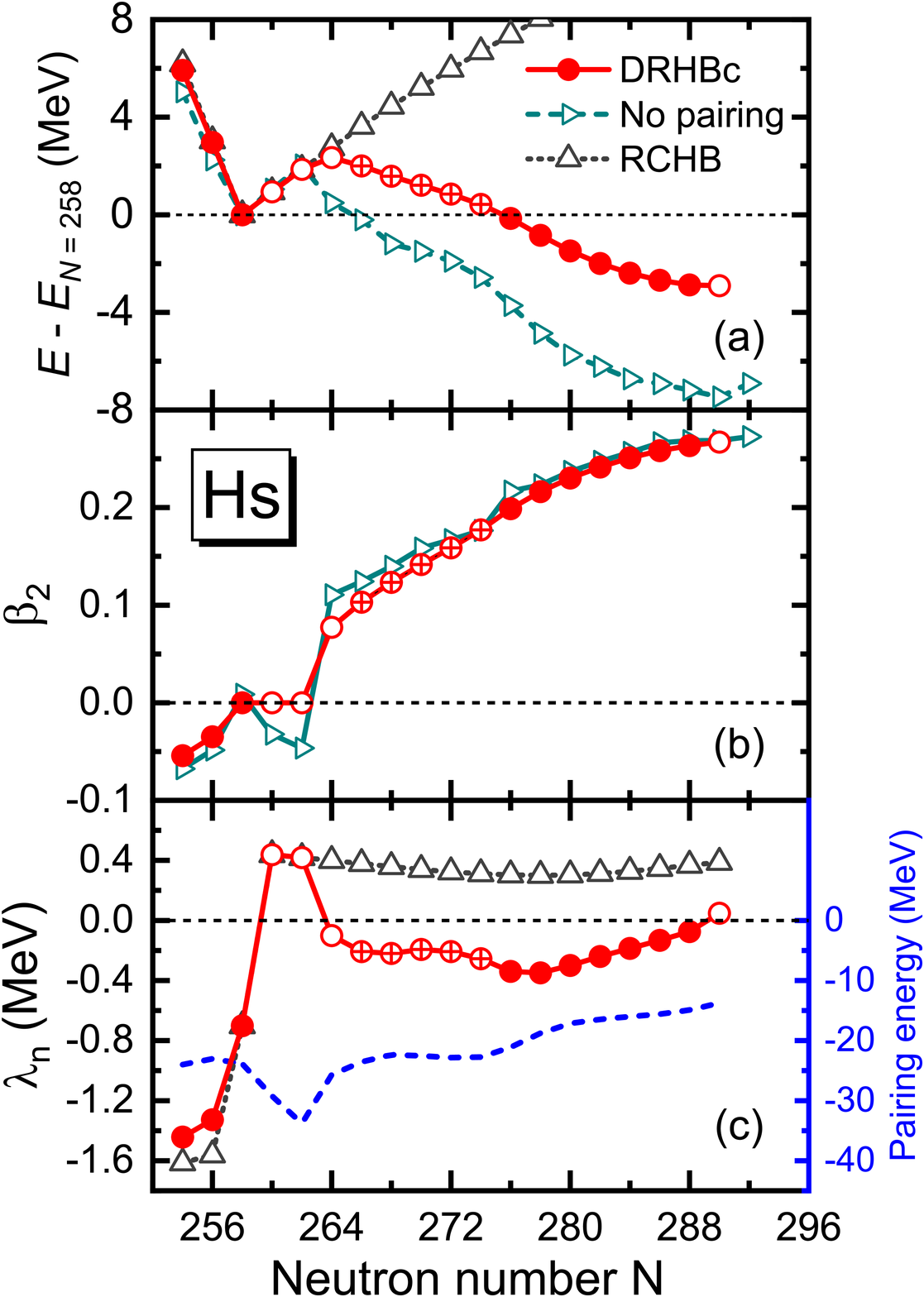}
  \caption{The DRHBc calculated (a) total energy relative to that of $^{366}$Hs ($N=258$), (b) quadrupole deformation $\beta_2$, and (c) neutron Fermi energy $\lambda_n$ as well as pairing energy, as functions of the neutron number $N$ for Hs isotopes near the two-neutron drip line. The results from the DRHBc without pairing and the RCHB calculations are shown for comparison. For the DRHBc results, the filled, open, and crossed symbols correspond to nuclei that are bound, unbound, and stable against two-neutron emission but unstable against multineutron emission, respectively.}
\label{fig3}
\end{figure}
%----------------------------------------------------------------------------------------

In order to study the underlying mechanism behind the emergence of the stability peninsula, taking $_{108}$Hs isotopes as examples, Fig.~\ref{fig3} shows the DRHBc calculated total energy relative to that of $^{366}$Hs$_{258}$, the quadrupole deformation, and the neutron Fermi energy as well as the pairing energy.
The DRHBc results without pairing and the RCHB results are also shown for comparison.
In Fig.~\ref{fig3}(a), there are three unbound, five two-neutron emission stable, and seven bound nuclei between the two-neutron drip line and the right border of the stability peninsula.
In Fig.~\ref{fig3}(b), the ground-state deformations for nuclei with $N=258,260,$ and $262$ are spherical and become prolate afterwards.
In Fig.~\ref{fig3}(c), the neutron Fermi energies for nuclei with $N=260$ and $262$ are positive and become negative at $N=264$, coinciding with the onset of deformation.
Although $^{372}$Hs$_{264}$ with a negative neutron Fermi energy gains binding energy from the deformation effects, it is shown in Fig.~\ref{fig3}(c) that a loss of pairing energy counteracts this, explaining the reason why $^{372}$Hs is still unstable against two-neutron emission.
After $N=264$, the ground-state deformation continues to increase and the stabilities against two- and multineutron emission emerge.
In the RCHB results with spherical symmetry, the total energy progressively increases and the neutron Fermi energy remains positive beyond $N=258$.
Therefore, the deformation effects play an essential role in the appearance of stabilities against two- and multineutron emission beyond the two-neutron drip line.
In the DRHBc results without pairing, it can be seen in Fig.~\ref{fig3}(b) that the absolute values of ground-state deformation are larger than those with pairing.
Starting at $N=264$, the nuclear system simply gains binding energy from the deformation effects, and thus the regions of stability against two- and multineutron emission are slightly changed.

%----------------------------------------------------------------------------------------
\begin{figure*}[htbp]
  \centering
  \includegraphics[width=0.8\textwidth]{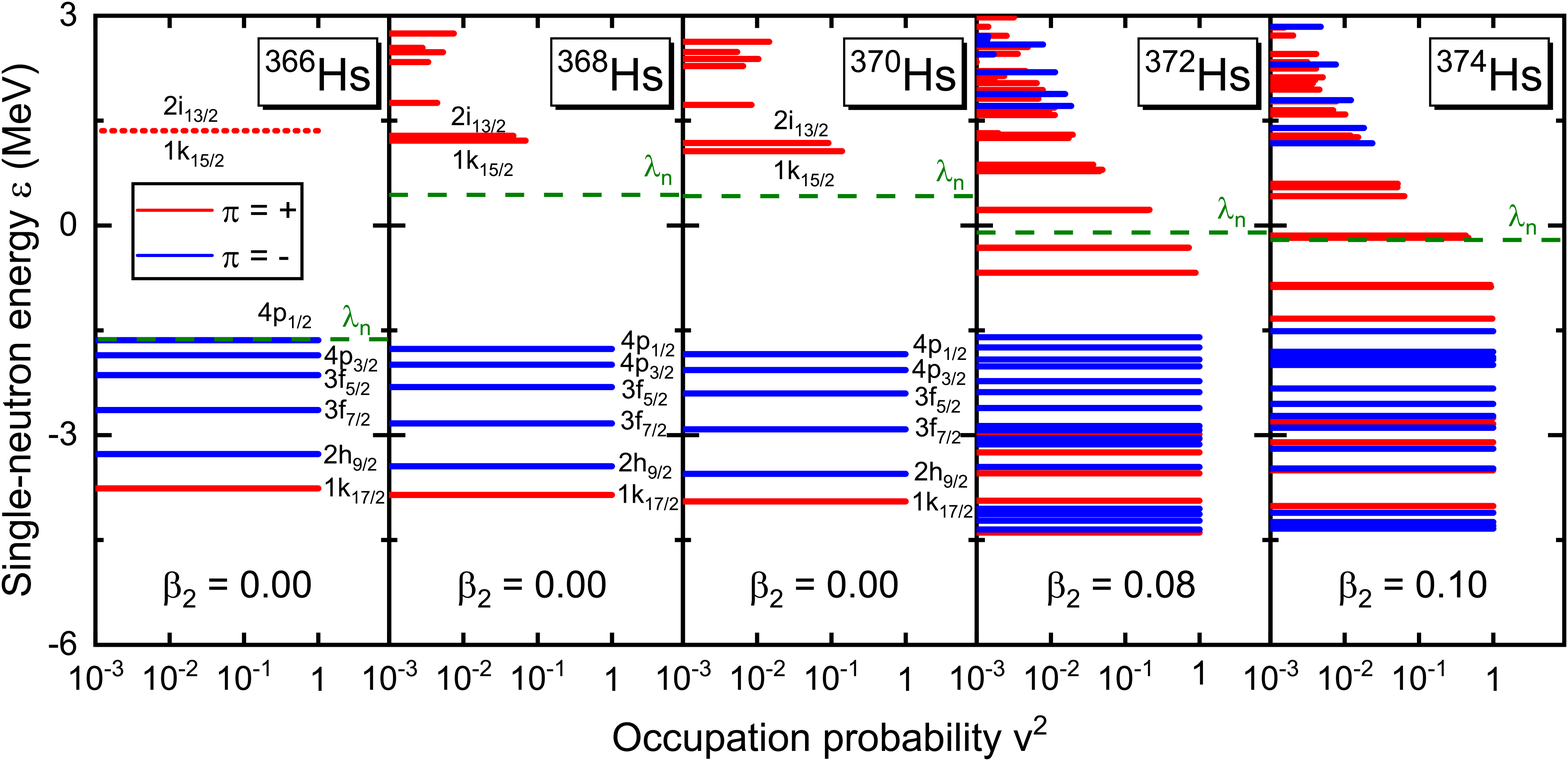}
  \caption{Single-neutron energy in the canonical basis around the Fermi energy versus the occupation probability $v^2$ for $^{366,368,370,372,374}$Hs in the DRHBc calculations. For spherical nuclei $^{366,368,370}$Hs, levels with $v^2\gtrsim 10^{-1}$ are labeled by their quantum numbers $nlj$. For $^{366}$Hs, the unoccupied levels $1k_{15/2}$ and $2i_{13/2}$ are shown by dotted lines.}
\label{fig4}
\end{figure*}
%----------------------------------------------------------------------------------------

To further explore the role of deformation effects in a microscopic way, the single-neutron spectra around the Fermi energy for $^{366,368,370,372,374}$Hs are shown in Fig.~\ref{fig4}.
One can see a clear gap about $3$ MeV between the occupied and the unoccupied levels in $^{366}$Hs, indicating a shell closure at $N=258$.
With more neutrons added to $^{366}$Hs, although the energies of single-neutron levels around the Fermi energy decrease slightly, the valence neutrons in $^{368}$Hs and $^{370}$Hs mainly occupy the $1k_{15/2}$ and $2i_{13/2}$ resonant orbits embedded in continuum, such that the two isotopes are unbound with positive Fermi energies.
Adding two more neutrons, $^{372}$Hs becomes deformed with $\beta_2 = 0.08$, and each $nlj$ orbit is split into $2j+1$ ones.
Two positive-parity orbits cross the continuum threshold and the neutron Fermi energy becomes negative for $^{372}$Hs.
However, as discussed in Fig.~\ref{fig3}, $^{372}$Hs is still unstable against two-neutron emission due to the loss of pairing energy.
For $^{374}$Hs, it becomes more deformed with $\beta_2 = 0.10$, one more orbit crosses the continuum threshold, and it becomes stable against two-neutron emission.
Hence, the deformation significantly influences the shell structure and results in stability against neutron emission beyond the two-neutron drip line.
Noteworthy here is another important deformation effect which results in a shallow connecting the known continent of stable nuclei and the predicted ``island of stability"; for details see Ref.~\cite{Meng2019SciChina} and references therein.

%----------------------------------------------------------------------------------------
\begin{figure*}[htbp]
  \centering
  \includegraphics[width=0.8\textwidth]{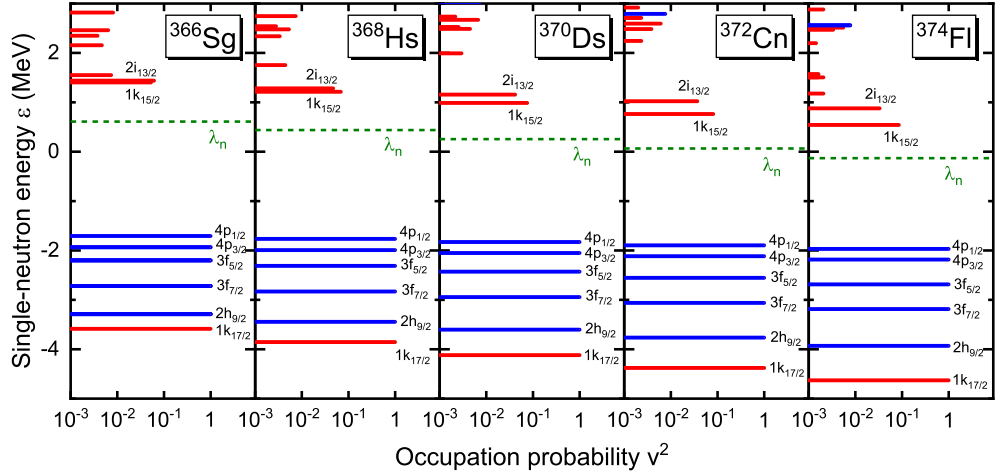}
  \caption{Same as Fig.~\ref{fig4}, but for $N=260$ isotones $^{366}$Sg, $^{368}$Hs, $^{370}$Ds, $^{372}$Cn, and $^{374}$Fl.}
\label{fig5}
\end{figure*}
%----------------------------------------------------------------------------------------

In Fig.~\ref{fig2}, it is shown that the appearance of stabilities against two- and multineutron emission is also relevant to $Z$.
For $N=260, 262$, and $264$ isotonic chains, nuclei with $Z\le 112$ are unbound and those with $Z\ge 114$ are bound.
To explore the underlying mechanism, the single-neutron spectra around the Fermi energy for the $N=260$ isotones $^{366}$Sg, $^{368}$Hs, $^{370}$Ds, $^{372}$Cn, and $^{374}$Fl are shown in Fig.~\ref{fig5}.
Due to the neutron shell closure at $N=258$, all these isotones exhibit a spherical shape.
The energies of single-neutron levels around the Fermi energy slightly decrease with increasing $Z$, similar to the case in Fig.~\ref{fig4} for $^{366,368,370}$Hs.
As a result, the neutron Fermi energy also decreases and becomes negative at $^{374}$Fl, making $^{374}$Fl bound.
It should be pointed out that, although the $1k_{15/2}$ and $2i_{13/2}$ orbits are located in continuum, their density distributions are localized in $^{374}$Fl due to the negative neutron Fermi energy~\cite{Dobaczewski1996PRC}.
For the weakly bound nuclei, the proper treatment of pairing correlations is of crucial importance~\cite{Dobaczewski1984NPA,Meng1998NPA}.
The DRHBc calculations without pairing show that the $N=260$ isotone will remain unbound until $Z=120$.
Therefore, the pairing correlations and continuum contribute to the stability peninsula.

%%%%%%%%%%%%%%%%%%%%%%%%%%%%%%%%%%%%%%%%%%%%%%%%%%%%%%%%%%
%                    begin  summary
%%%%%%%%%%%%%%%%%%%%%%%%%%%%%%%%%%%%%%%%%%%%%%%%%%%%%%%%%%

%----------------------------------------------------------------------------------------
%\section{Summary}\label{summary}
%----------------------------------------------------------------------------------------

In summary, the predictive power of the deformed relativistic Hartree-Bogoliubov theory in continuum for nuclear mass in the superheavy region, $102 \le Z \le 120$, has been examined.
The accuracy to predict the 10 (56) measured (measured and empirical) masses is $0.635$ ($0.642$) MeV, which is in contrast to the $0.515$ ($1.360$) MeV by WS4 and $0.910$ ($2.831$) MeV by FRDM.

The stabilities against two- and multineutron emission beyond the two-neutron drip line are explored by the DRHBc theory, which takes into account simultaneously the deformation effects, the pairing correlations, and the continuum effects.
Nuclei stable against two- or multineutron emission beyond the two-neutron drip line are predicted in the $_{106}$Sg, $_{108}$Hs, $_{110}$Ds, and $_{112}$Cn isotopic chains, forming a peninsula of stability adjacent to the nuclear mainland.
To explore the underlying mechanism, by taking Hs isotopes near the drip line as examples, the total energy, neutron Fermi energy, and ground-state deformation are studied and compared with the results in the RCHB and DRHBc without pairing calculation.
The single-neutron spectra around the Fermi surface for several Hs isotopes and $N = 260$ isotones are analyzed.
The formation of the stability peninsula beyond the two-neutron drip line is essentially due to the deformation which significantly affects the shell structure around the Fermi surface.
The pairing correlations and continuum influence the stability peninsula in a self-consistent way.

It would be interesting to study the stability against two- and multineutron emissions beyond the two-neutron drip line in other mass regions and the deformation effects of higher orders.
Work along this line is in progress~\cite{Pan2021arXiv,He2021arXiv}.

Finally, it is interesting to discuss the stabilities against other decay modes, such as nuclear fission and $\beta^-$ decay, for the nuclei stable against neutron emission beyond the neutron drip line.
To examine the stability against fission, taking $^{384}$Hs and $^{396}$Hs as examples, constrained DRHBc calculations have been performed.
Their fission barrier heights are found to be  around $2.0$ and $4.6$ MeV, respectively.
The fission half-life is sensitive to not only the barrier height but also other quantities such as the width and shape of the barrier.
Based on the systematics on the half-life versus the fission barrier height in Ref.~\cite{Smolanczuk1995PRC}, the fission half-life for $^{384}$Hs ranges from femtoseconds to seconds, and that for $^{396}$Hs ranges from picoseconds to minutes.
To examine the stability against $\beta^-$ decay, taking $^{384\sim396}$Hs as examples, their $Q_{\beta^-}$ values are calculated to be around $12-15$ MeV.
According to the Sargent law~\cite{Sargent1933}, these $Q_{\beta^-}$ values correspond to $\beta^-$ decay half-lives of the order of 10 ms.
Future studies on the stability of these very exotic nuclei with triaxial deformation and other higher multipoles are expected.

%\begin{acknowledgments}

Helpful discussions with members of the DRHBc Mass Table Collaboration are gratefully appreciated. This work was partly supported by the National Natural Science Foundation of China (Grants No.~11935003, No.~11875075, No.~11975031, No.~12070131001, No.~11775112, and No.~U2032138), the National Key R\&D Program of China (Contracts No.~2017YFE0116700 and No.~2018YFA0404400), the State Key Laboratory of Nuclear Physics and Technology, Peking University (Grant No.~NPT2020ZZ01), and High-performance Computing Platform of Peking University.

%\end{acknowledgments}

%======================================================================================%

%\end{CJK*}
\end{document}